\documentstyle[12pt]{article}
\newcommand{\tr}{{\rm tr\,}}
\newcommand{\cotan}{{\rm cotan\,}}
\newcommand{\sss}{\scriptscriptstyle}

\def\ba{\begin{array}}
\def\ea{\end{array}}
\def\beq{\begin{eqnarray*}}
\def\eeq{\end{eqnarray*}}
 
\begin{document}
\begin{center}
{\Large \bf Absence of Axial Anomaly in the Background of the Bohm-Aharonov 
Vector Potential}

\bigskip
{\large Yu.A.~Sitenko}
\date{}
\smallskip

{\it
Bogolyubov Institute for Theoretical Physics,\\
National Academy of Sciences,
03143 Kyiv, Ukraine
}
\end{center}

\begin{abstract}

The problem of the axial anomaly in the presence of the Bohm-Aharonov gauge 
vector field is exactly solved.            

\end{abstract}

The axial anomaly arises as a violation of the classical conservation law 
for the axial-current at the quantum level. Since its discovery \cite{Schw,Adl,Bel} 
the anomaly has played a more and more significant role in the 
development of contemporary quantum field theory, and has led to a number of
important phenomenological consequences in particle physics. Although the first 
anomalies were found in studies by means of the diagram technique of 
perturbations in coupling constant, it was soon recognized that the 
results do not depend on the use of perturbative methods. The nonperturbative 
(i.e., not describable in 
the framework of perturbation theory) nature of the anomalies can be revealed 
by means of an approach in which the gauge vector field is treated as a 
classical external one and the problem of quantizing massless fermions in 
this background is solved. Such a treatment makes it possible to regard the 
anomaly as a manifestation of nontrivial topology of configurations
of the gauge vector field and establish a connection between the anomaly 
and the topological invariant of the spectrum of the massless Dirac operator in 
an external-field background.

Singular (or contact or zero-range) interaction potentials were
introduced in quantum mechanics more than sixty years ago \cite{Bet,Tho,Fer}.
A mathematically consistent and rigorous treatment of the subject
was developed \cite{Ber}, basing on the notion of self-adjoint extension
of a Hermitian operator (for a review see monograph
\cite{Alb}). Singular external-field background can act on the quantized 
spinor field in a rather unusual manner: a leak of quantum numbers from the 
singularity point into the vacuum occurs \cite{Gro,Yam,Sit96,Sit97,Sit99,Sit9}. 
This is due to the fact that a solution to the Dirac 
equation, unlike that to the Schrodinger one, does not obey a condition of 
regularity at the singularity point. It is necessary then to specify a boundary 
condition at this point, and the least restrictive, but still physically 
acceptable, condition is such that guarantees self-adjointness of the 
operator of the appropriate dynamical variable.

In the present paper the problem of the axial anomaly in the singular 
background of the Bohm-Aharonov \cite{Aha} gauge vector field is 
comprehensively studied. We show 
that, contrary to the leak of vacuum quantum numbers, the leak of anomaly from 
the singularity point does not occur.


Let us consider the effective action functional for quantized massless spinor 
field $\Psi(x)$ in external classical vector field $V_\mu (x)$ in the 
Wick-rotated (Euclidean) $d$-dimensional space-time
\begin{equation}
S^{\rm eff}[V_\mu (x)] = -\ln  \int d\Psi(x) \,d\Psi^\dagger(x)\,
\exp[-\int d^d x \,L(x)] = -\ln Det (-i\gamma^\mu\nabla_\mu) ,
\end{equation}
where 
\begin{equation}
L(x) = -{i\over2} \Psi^\dagger(x)\gamma^\mu [\nabla_\mu \Psi(x)] +
{i\over2} [\nabla_\mu \Psi(x)]^\dagger \gamma^\mu \Psi(x)
\end{equation}
is the Lagrangian density, 
$\nabla_\mu = \partial_\mu-iV_\mu (x)$
is the covariant differentiation operator, and $\gamma^\mu$     
$(\mu=\overline{1,d})$ are the Dirac matrices, 
\begin{equation}
[\gamma^\mu,\gamma^\nu]_+ = 2g^{\mu\nu}, \qquad \tr\gamma^\mu = 0,  \qquad
g_{\mu\nu} = diag (1, ... ,1) .
\end{equation}
If there exists matrix $\Gamma$ anticommuting with the Dirac matrices,
\begin{equation}
[\Gamma,\gamma^\mu]_+=0 ,\qquad  \tr\Gamma=0 ,\qquad  \Gamma^2=I ,
\end{equation}
then one can define local chiral transformation
\begin{eqnarray}
\Psi(x) \rightarrow e^{i\omega(x)\Gamma} \Psi(x), \qquad 
\Psi^\dagger(x) \rightarrow \Psi^\dagger(x) e^{i\omega(x)\Gamma},
\nonumber \\
V_\mu (x) \rightarrow e^{i\omega(x)\Gamma} V_\mu (x) e^{-i\omega(x)\Gamma} + 
\partial_\mu \omega(x)\Gamma .
\end{eqnarray}
The invariance of functional (1) under this transformation corresponds to 
conservation law
\begin{equation}
\nabla_\mu J^\mu_{d+1} (x) =0,
\end{equation}
where
\begin{equation}
J^\mu_{d+1} (x) = i\tr\langle x| \gamma^\mu \Gamma 
(-i\gamma^\nu \nabla_\nu)^{-1}| x\rangle.
\end{equation} 
However, functional (1), as well as current (7), is ill-defined, 
suffering from both ultraviolet and infrared divergences. Performing the 
regularization of divergencies in a way which is consistent with gauge 
invariance, one may arrive at the violation 
of conservation law (6) (i.e. at the axial anomaly) \cite{Schw,Adl,Bel}. 

An example of a singular background field configuration is 
provided by that of the Bohm-Aharonov \cite{Aha} vortex represented 
by a point for $d = 2$, a line for $d = 3$,
and a $(d - 2)$-dimensional hypersurface for $d > 3$:
\begin{equation}
V^1(x)=-\Phi^{(0)} {x^2\over (x^1)^2+(x^2)^2}, \quad V^2(x)=\Phi^{(0)} 
{x^1\over (x^1)^2+(x^2)^2}, \quad  V^\nu(x)=0, \,\,\, \nu=\overline{3,d},
\end{equation}
\begin{equation}
B^{3 \cdot\cdot\cdot d}(x) = 2\pi\Phi^{(0)} \delta(x),
\end{equation}
where $\Phi^{(0)}$ is the vortex flux in $2\pi$  units, i.e.
in the London ($2\pi \hbar c e^{-1}$) units, since we use
conventional units $\hbar=c=1$ and coupling constant $e$ is included into
vector potential $V_\mu (x)$.

In the $d=2$ case, the $\gamma$-matrices are chosen as
$\gamma^1 = \sigma_1 ,    \gamma^2 = \sigma_2 $,
and, consequently,
$\Gamma = \sigma_3$ ,
where $\sigma_1,\sigma_2$ and $\sigma_3$ are the Pauli matrices. Then the 
complete set of solutions to Dirac equation
\begin{equation}
(-i\gamma^\mu\nabla_\mu - E)\langle x|E \rangle = 0
\end{equation}
in background (8) takes form
\begin{equation}
\langle x|E \rangle = \sum_{n\in{\rm Z}}\left(\begin{array}{l}f_n(r)
\exp(i n \varphi) \\g_n(r)\exp[i(n+1)\varphi]\end{array}\right),
\end{equation}
where $\rm Z$ is the set of integer numbers,
 $r$ and  $\varphi$ are the polar coordinates,
and the radial functions, in general, are
\begin{equation}
\left(\begin{array}{l}f_n(r)\\g_n(r)\end{array}\right)=
\left(\begin{array}{c}C^{(1)}_n(E)J_{n-\Phi^{(0)}}(|E|r)+
C^{(2)}_n(E)Y_{n-\Phi^{(0)}}(|E|r)\\
i ({E/|E|})\bigl[C^{(1)}_n(E)J_{n+1-\Phi^{(0)}}(|E|r)+
C^{(2)}_n(E)Y_{n+1-\Phi^{(0)}}(|E|r)\bigr]\end{array}\right),
\end{equation}
$J_\rho(u)$ and $Y_\rho(u)$ are the Bessel and the
Neumann functions of order $\rho$. It is clear that the condition
of regularity at $r=0$ can be imposed on both $f_n$ and $g_n$ for all
$n$ in the case of integer values of quantity
$\Phi^{(0)}$ only. Otherwise, the condition of regularity at
$r=0$ can be imposed on both $f_n$ and $g_n$ for all but $n=n_0$,
where $n_0$ is the integer part of the quantity $\Phi^{(0)}$ (i.e. the
integer which is less than or equal to $\Phi^{(0)}$); in this case at least one
of the functions, $f_{n_0}$ or $g_{n_0}$, remains irregular, although
square integrable, with the asymptotics $r^{-p}$ $(p<1)$ at $r \to 0$.  
The question arises then, what boundary condition, instead of
regularity, is to be imposed on $f_{n_0}$ and $g_{n_0}$ at $r=0$ in
the latter case?

To answer this question, one has to find the self-adjoint extension
for the partial Dirac operator corresponding to the mode with $n=n_0$.
If this operator is defined on the domain of regular at $r=0$
functions, then it is Hermitian, but not self-adjoint, having the
deficiency index equal to (1,1).  
The use of the Weyl - von Neumann theory 
of self-adjoint operators (see, e.g., Ref.\cite{Alb}) yields that, for the 
partial Dirac operator to be self-adjoint extended, it has to be defined 
on the domain of functions satisfying the boundary condition 
\begin{eqnarray}
i\cos\biggl({\theta\over2}+{\pi\over4}\biggr)\,2^{1-F}\,\Gamma (1-F)\,
\lim_{r\to0}\biggl(\mu r\biggr)^Ff_{n_0}(r) 
\nonumber \\
=\sin\biggl({\theta\over2}
+{\pi\over4}\biggr)\,2^F\,\Gamma (F)\,\lim_{r\to0}
\biggl(\mu r\biggr)^{1-F}g_{n_0}(r),
\end{eqnarray}
where $\Gamma(u)$ is the Euler gamma function,
\begin{equation}
F=\Phi^{(0)}-n_0
\end{equation}
is the fractional part of quantity $\Phi^{(0)}$  ($0\leq F <1$), $\theta$ is 
the self-adjoint extension parameter, and $\mu>0$ is inserted merely for 
the dimension reasons. Note that  Eq.(13) 
implies that $0<F<1$, since in the case of $F=0$ both $f_{n_0}$ and $g_{n_0}$ 
satisfy the condition of regularity at $r=0$. Note also that, since Eq.(13) 
is periodic in $\theta$ with period $2\pi$, all permissible values of 
$\theta$ can be restricted, without a loss of generality, to range
$0 \leq \theta \leq 2\pi$.


The gauge invariant regularization of $\nabla_\mu J^\mu_{d+1} (x)$ can be 
achieved by means of the zeta function method \cite{Sal,Dow,Hawk}, yielding, 
instead of Eq.(6), the following relation 
\begin{equation}
\nabla_\mu J^\mu_{d+1} (x) = 2 \,\lim_{z\rightarrow 0} \,
\lim_{M\rightarrow 0} \,\,
\tilde{\zeta}_x(z|M),
\end{equation}
where
\begin{equation}
\tilde{\zeta}_x(z|M)=\tr\langle x|\Gamma\,\bigl\{\nabla^\mu \nabla_\mu + 
{i\over 2}[\gamma^\mu,\gamma^\nu]_{-}[\nabla_\mu V_\nu (x)] + M^2 \bigr\}^{-z}
|x \rangle
\end{equation}
is the modified zeta function density.

In the $d=2$ case, using the explicit form of the solution to the Dirac 
equation in background (8), it is straightforward to compute the modified 
zeta function density. As follows already from the preceding discussion, 
the modified zeta function density vanishes in the case of integer values of 
$\Phi^{(0)}$ $(F=0)$, since this case is indistinguishable from the case of 
the trivial background ($\Phi^{(0)}=0$). In the case of noninteger values
of $\Phi^{(0)}$ ($0<F<1$)
we get 
\begin{eqnarray}
\tilde{\zeta}_x(z|M)={\sin(F\pi)\over\pi^3}
\sin(z\pi)r^{2(z-1)}\int\limits_{|M|r}^\infty dw\, w(w^2-M^2r^2)^{-z}
\biggl\{K_F^2(w)-K_{1-F}^2(w)
\nonumber \\
+ \bigl[K_F^2(w)+K_{1-F}^2(w)\bigr]
\tanh\ln\bigl[\bigl({w\over \mu r}\bigr)^{2F-1} 
\cotan \bigl({\theta\over2}-{\pi\over4}\bigr)\bigr]\biggr\}, 
\end{eqnarray}
Taking limit $M\rightarrow 0$, we get 
\begin{eqnarray}
\tilde{\zeta}_x(z|0)={\sin(F\pi)\over\pi^3}\sin(z\pi)r^{2(z-1)}\biggl\{
{\sqrt{\pi}\over2}\, {\Gamma(1-z)\over\Gamma({3\over2}-z)}
\bigl(F-{1\over2})\Gamma(F-z)\Gamma(1-F-z)
\nonumber \\
+\int\limits_0^\infty dw\, w^{1-2z}\bigl[K_F^2(w)+K_{1-F}^2(w)\bigr]
\tanh\ln\bigl[\bigl({w\over \mu r}\bigr)^{2F-1} 
\cotan \bigl({\theta\over2}-{\pi\over4}\bigr)\bigr]\biggr\};
\end{eqnarray}
in particular, at half-integer values of the vortex flux:
\begin{equation}
\tilde{\zeta}_x(z|0)\big|_{F={1\over2}}={\sin\theta\over
2\pi^{\sss3\over\sss2}}\, {\Gamma({1\over2}-z)\over\Gamma(z)} r^{2(z-1)};
\end{equation}
and at $\cos\theta=0$:
\begin{equation}
\tilde{\zeta}_x(z|0)=\pm{\sin(F\pi)\over2\pi^{\sss3\over\sss2}}\,
{\Gamma({3\over2}-z\pm F\mp{1\over2})\Gamma({1\over2}-z\mp
F\pm{1\over2})\over \Gamma(z)\Gamma({3\over2}-z)}\, r^{2(z-1)}, \quad
\theta=\pi (1 \mp {1\over2}).
\end{equation}
Consequently, we obtain
\begin{equation}
\tilde{\zeta}_x(0|0)=0, \qquad x\neq0. 
\end{equation}

Thus the anomaly is absent everywhere on the plane with the puncture at
$x=0$. This looks rather natural, since twodimensional anomaly
$2\tilde{\zeta}_x(0|0)$ is usually identified with
quantity ${1\over\pi} B(x)$,
and background field strength $B(x)$ vanishes everywhere on the
punctured plane, see Eq.(9) at $d=2$. We see that natural anticipations
are confirmed, provided that the boundary conditions at the puncture
are chosen to be physically acceptable, i.e., compatible with the
self-adjointness of the Dirac operator.
        

We might finish here the discussion of the anomaly problem in the
background of the Bohm-Aharonov vortex. However, there remains a
purely academic question: what is the anomaly in background
(8)-(9) on the whole plane (without puncturing
$x=0$)? Just due to a confusion persisting in the literature
\cite{Sold,Mor}, we shall waste now some time to clarify this, otherwise
inessential, point.

Background field strength (9), when considered on the whole
plane, is interpreted in the sense of a distribution (generalized
function), i.e., a functional on a set of suitable test functions
$f(x)$:
\begin{equation}
\int d^2x\,f(x)\,{1\over\pi}B(x)=f(0)\, 2 \Phi^{(0)} ; 
\end{equation}
here $f(x)$ is a continuous function. In particular, choosing
$f(x)=1$, one gets
\begin{equation}
\int d^2x\,{1\over\pi}B(x)=2 \Phi^{(0)}. 
\end{equation}
Considering the anomaly on the whole plane, one is led to study
different limiting procedures as $r\rightarrow 0$ and $z\rightarrow 0$
in Eq.(18). So, the notorious question is, whether anomaly
$2\tilde{\zeta}_x$ can be interpreted in the sense of a
distribution which coincides with distribution
${1\over\pi}B(x)$? The answer is resolutely negative, and
this will be immediately demonstrated below.

First, using explicit form (18), we get
\begin{equation}
\int d^2x\,2\tilde{\zeta}_x(z|0) = 
\left\{\ba{cc} \infty,  &z\neq0\\ 0,  &z=0\\ \ea \right.\, ;
\end{equation}
therefore, the anomaly functional cannot be defined on the same set of
test functions as that used in Eq.(22) (for example, the test
functions have to decrease rapidly enough at large (small) distances in
the case of $z>0$ ($z<0$)). Moreover, if one neglects the requirement
of self-consistency, permitting a different (more specified) set of test
functions for the anomaly functional, then even this will not save the
situation. Let us take $z>0$ for definiteness and use the test functions
which are adjusted in such a way that the quantity
\begin{equation}
A=\lim_{z\rightarrow 0_{+}}\int d^2x\,f(x)\,2\tilde{\zeta}_x(z|0)
\end{equation}
is finite. Certainly, this quantity can take values in a rather wide
range, but it cannot be made equal to the right-hand side of Eq.(23).
Really, the only source of the dependence on $\Phi^{(0)}$ in the
integral in Eq.(25) is the factor $\tilde{\zeta}_x(z|0)$, and the
latter, as is evident from Eq.(18), depends rather on $F$,
than on $\Phi^{(0)}$ itself, thus forbidding the linear dependence of $A$ on
$\Phi^{(0)}$. In particular, let us choose test function $f(x)$ in the form
\begin{equation}
f(x)=\exp(-{\tilde{\mu}}^{2}\,r^{2}), 
\end{equation}
where $\tilde{\mu}$ is the parameter of the dimension of mass. Then, choosing
the case of $\cos\theta=0$ for simplicity and using Eq.(20), one gets that
Eq.(25) takes form
\begin{equation}
A=2\,\bigl(F-{1\over2}\pm{1\over2}\bigr),
\qquad \theta=\pi (1 \mp {1\over2}), 
\end{equation}
which differs clearly from $2\Phi^{(0)}$.


We have proved that, in a singular background, the conventional relation 
between the axial anomaly and the background field strength is valid only in
the space with punctured singularities; consequently, wherever the field 
strength is zero the anomaly always is absent. If singularities are 
not punctured, then the anomaly and the field strength can be interpreted 
in the sense of distributions, but, contrary to the assertions of the 
authors of Refs.\cite{Sold,Mor}, the conventional relation is not valid.

\vfill\eject

\end{document}